\documentclass[runningheads]{cl2emult}

\usepackage{makeidx}  
\usepackage{graphicx} 
\usepackage{subeqnar} 
\usepackage{multicol} 
\usepackage{cropmark} 
\usepackage{subfigure}
\usepackage{epsfig}
\makeindex            

\begin{document}

%
\title*{Microscopic Models of Financial Markets\protect\newline}
\toctitle{Microscopic Models of Financial Markets}
%
%
\titlerunning{Microscopic Models of Financial Markets}
%
\author{E. Samanidou\inst{1}
\and E. Zschischang\inst{1} 
\and D. Stauffer\inst{2}
\and T. Lux\inst{1}}
\authorrunning{E. Samanidou, E. Zschischang {\em et al.}}
%
%
\institute{Department of Economics, University of Kiel, Olshausenstrasse 40, D-24118 Kiel, Euroland
\and Department of Physics, University of Cologne, D-50923 Cologne, Euroland}

\maketitle              
%

\begin{abstract} \label{abstract}
This review deals with several microscopic models of financial markets which have been studied by economists and physicists over the last decade: Kim-Markowitz, Levy-Levy-Solomon, Cont-Bouchaud, Solomon-Weisbuch, Lux-Marchesi, Donangelo-Sneppen and Solomon-Levy-Huang. After an overview of simulation approaches in financial economics, we first give a summary of the Donangelo-Sneppen model of monetary exchange and compare it with related models in economics literature. Our selective review then outlines the main ingredients of some influential early models of multi-agent dynamics in financial markets (Kim-Markowitz, Levy-Levy-Solomon). As will be seen, these contributions draw their inspiration from the complex appearance of investors' interactions in real-life markets. Their main aim is to reproduce (and, thereby, provide possible explanations) for the spectacular bubbles and crashes seen in certain historical episodes, but they lack (like almost all the work before 1998 or so) a perspective in terms of the universal statistical features of financial time series. In fact, awareness of a set of such regularities (power-law tails of the  distribution of returns, temporal scaling of volatility) only gradually appeared over the nineties. With the more precise description of the formerly relatively vague characteristics (e.g. moving from the notion of fat tails to the more concrete one of a power-law with index around three), it became clear that financial markets dynamics give rise to some kind of universal scaling laws. Showing similarities with scaling laws for other systems with many interacting sub-units, an exploration of financial markets as multi-agent systems appeared to be a natural consequence. This topic was pursued by quite a number of contributions appearing in both the physics and economics literature since the late nineties. From the wealth of different flavors of multi-agent models that have appeared by now, we discuss the Cont-Bouchaud, Solomon-Levy-Huang and Lux-Marchesi models. Open research questions are discussed in our concluding section.  
\end{abstract}
\bigskip
Submitted to F. Schweitzer (ed.),  Microscopic Models for Economic Dynamics,
 Lecture notes in physics, Springer, Berlin-Heidelberg 2002.
\end{document}